# Which country epitomizes the world?
# A study from the perspective of demographic composition

**Takahiro Yoshida, Ph.D.**

*Center for Global Environmental Research, National Institute for Environmental Studies,*
*16-2 Onogawa, Tsukuba, Ibaraki 305-8506, Japan.*
*E-mail: yoshida.takahiro@nies.go.jp (corresponding author)*

**Rim Er-Rbib**

*Graduate School of Systems and Information Engineering, University of Tsukuba,*
*1-1-1 Tennodai, Tsukuba, Ibaraki 305-8573, Japan.*
*E-mail: s1720532@u.tsukuba.ac.jp*

**Morito Tsutsumi, Ph.D.**

*Faculty of Engineering, Information and Systems, University of Tsukuba,*
*1-1-1 Tennodai, Tsukuba, Ibaraki 305-8573, Japan.*
*E-mail: tsutsumi@sk.tsukuba.ac.jp*

*Acknowledgements:*
We received helpful comments from participants at the 7th International Workshop on Compositional Data Analysis (Siena, Italy; June, 2017). We would particularly like to thank them. This study was supported by the Japan Society for the Promotion of Science under two Grants-in-Aid for Scientific Research 16J02219 and 17K18554.

# Title:

Which country epitomizes the world?
A study from the perspective of demographic composition

# Abstract:

Demographic indicators are an essential element in considering various problems in the social economy, such as predicting economic fluctuations and establishing policies. The literature widely discusses the growth of the world population or issues pertaining to its aging, but has given little to no attention to population structures and transition patterns. In this article, we take advantage of the characteristics of compositional data to examine the transition of the world population structure. Using the Aitchison distance, we examine the similarity of the world population structure from the 1990s to 2080 and that of countries and regions in 2015 and create maps to illustrate the results. Accordingly, we identify the following countries and regions as epitomes of the world's population structure through different periods: India, Northern Africa and South Africa, in the 1990s, South America in 2015 to 2030, Oceania and Northern America in 2040, Uruguay and Puerto Rico in 2050 to 2060, and Italy and Japan in the distant future. We then cluster countries based on the similarity of their population structures in 2015 and correspond each cluster to a certain period. We found that Russia and Western Europe gather in a cluster that does not correspond to any period, indicating a recessive population structure.

# 1. Introduction

In the past century, demographers focused on the total size of the population and fertility rate. However, today, as May (2012) points out in his book "World Population Policies," their main concern is the composition of the population, namely age groups and their interrelations. Examining the shape of the population pyramid helps to clarify the current demographic situation, excavate past events, and forecast future trends (e.g., Bivand et al., 2017). Moreover, advances made in the Compositional Data Analysis (CoDA) (Aitchison, 1986; Aitchison and Egozcue, 2005; Pawlowsky-Glahn et al., 2015), which we describe later, enable a comprehensive examination of population structure without focusing on a particular age group or the total size.

When considering international policies and global visions, analysis at the aggregate level is necessary. As such, it is important to consider the world population structure as a decisive and highly influential element in global progress. A holistic vision of how the world population will shift through the decades can hint towards the dominant demographic trend at different periods, and thus provide a roadmap by which international organizations and world leaders can design international policies and set sustainable development goals.

In this article, we create maps that showcase the countries that epitomize the world population through time, illustrating how the world population structure has changed since 1990 and how it will transition in the next half century. Rather than examining each age group separately, we focus on population composition ratios and the relative transition of age groups.

To answer this question, we search for an epitome of the world by investigating the similarity of population structures between countries in 2015 and the world in different years (Figure 1). We first present the data and methodology employed. Then, based on the analysis, we showcase the countries that epitomize the world population structure from 1990 to 2080 and their characteristics. Finally, we discuss and summarize the results using a cluster analysis of the population on an international level.

**[Figure 1: Analysis scheme: Comparison between world population structures of countries and areas in 2015 and different points of time], around here**

## 2. Methods and Materials

Few population studies employ composition ratio data of the population. As far as the authors know, the studies by Loyd (2010; 2016a; 2016b) and Lloyd et al. (2012) are the only ones that focus on the spatial concentration of the population and target the population at the regional or national levels in the United Kingdom. In this article, we also take advantage of the characteristics of population composition ratio data to compare the population structure at global and international levels. This comparison and finding an epitome of the world can help identify the demographic characteristics of the world by examining the present characteristics of a country that epitomizes the world. Specifically, it is possible to determine which country or region currently has a population composition ratio similar to that of the world over time, as illustrated in Figure 1, by measuring the distance between population composition ratio data (vectors). To account for the characteristics of the compositional data (CoDa), we employ the Aitchison distance, as explained later. We calculate the Aitchison distance between the population ratios of each country in 2015 and the population ratios of the world from 1990 to 2080 every ten years. To summarize and create a comprehensive epitome of the world throughout all years, we performed hierarchical clustering based on the similarity between the population structures of countries in 2015 using Aitchison's distance.

## 2.1. Compositional data analysis and Aitchison distance

The population composition is CoDa, which are multidimensional data consisting of ratios (Aitchison, 1986). A fundamental feature of CoDa is that each component is subject to a positive number constraint and constant sum constraint. For these two constraints, where $D$ is the rank, the sample space of CoDa differs from the real $D$-space $R^D$ and is limited to the simplex $S^D \subset R^D$, as represented by equation (1):

$$S^D = \left\{ \boldsymbol{x} = [x_1, x_2, \ldots, x_D]^t \middle| x_d > 0, d = 1, 2, \ldots, D; \sum_{d=1}^{D} x_D = k \right\}, \tag{1}$$

where superscript $t$ denotes an operator representing a transpose of a vector and $k$ is a constant, for example, $k = 1$ when the component is expressed as a ratio, and $k = 100$ when expressed as a percentage.

In the statistics field, CoDA refers to using ratio data such as that of population composition.

These population composition characteristics provide us with the benefit of using CoDA in population analysis when focusing on the population composition ratio, as it enables a comparison of population pyramids, regardless of the total population size.

We explain the distance index used to measure the similarity of population composition ratios among countries and regions. For CoDa, Aitchison's distance $d_{\text{Ait}}$ expressed in equation (2) is widely used:

$$d_{\text{Ait}}(\boldsymbol{x}, \boldsymbol{y}) = d_{\text{Euc}}(\text{clr}(\boldsymbol{x}), \text{clr}(\boldsymbol{y}))$$

$$= \sqrt{\sum_{d=1}^{D}\left(\ln\left(\frac{x_d}{g(\boldsymbol{x})}\right) - \ln\left(\frac{y_d}{g(\boldsymbol{y})}\right)\right)^2}$$

$$= \sqrt{\sum_{d=1}^{D}\left(\ln\left(\frac{x_d}{y_d} \cdot \frac{g(\boldsymbol{y})}{g(\boldsymbol{x})}\right)\right)^2} \quad (2)$$

where $\boldsymbol{x}, \boldsymbol{y} \in S^D$; $g(\boldsymbol{x}) = \left(\prod_{d=1}^{D} x_D\right)^{1/D}$, the geometric mean of $\boldsymbol{x}$ (it is the same for $\boldsymbol{y}$); and clr is the centered log-ratio transformation, a function defined from the simplex $S^D$ to the real space $R^D$, expressed as in equation (3):

$$\text{clr}(\boldsymbol{x}) = \left[\ln\left(\frac{x_1}{g(\boldsymbol{x})}\right), \ln\left(\frac{x_2}{g(\boldsymbol{x})}\right), \ldots, \ln\left(\frac{x_D}{g(\boldsymbol{x})}\right)\right]^t. \quad (3)$$

Euclid distance $d_{\text{Euc}}$ is an index focusing on the difference between values of components, while Aitchison's distance focuses on the difference of the relative values of the components, or, in other words, how many times is the difference between the values of the component (Aitchison et al., 2000; Otero, 2005; Lovell et al., 2011; Pawlowsky-Glahn et al., 2015; Seya and Yoshida, 2017 for examples on comparison between $d_{\text{Ait}}$ and $d_{\text{Euc}}$).

**2.2. Demographic composition**

The data used in this analysis were the population projections from the "Quinquennial Population by Five-Year Age Groups—Both Sexes. De facto population as of 1 July 2015 classified by five-year age groups (0–4, 5–9, 10–14, ..., 95–99, 100+)" (United Nations, 2015). The 2015 revision covers a 150-year time horizon subdivided into past estimates (1950–2015) and future projections (2015–2100). Population estimates and projections were carried out for 233 countries or areas. This article publishes the detailed results of only 201 countries or areas with 90,000 inhabitants or more in 2015[1]. Aggregated results by geographic region, the UN's country classification based on level of development, and World Bank's classification based on income (see United Nations, 2015) were also included in the analysis to obtain more insightful results.

# 3. Results and discussion

The purpose of this article was to identify the epitome of the world population throughout the years from the recent past (1990) to the distant future (2080) based on population composition ratios to study the characteristics of the global population and its transition. We calculate the Aitchison distance, and by examining the differences between the shapes of the population pyramids, only pick up countries for which the Aitchison distance is smaller than one, as the shapes of the pyramids are not similar when it exceeds one (Figure 2). We focus only on the medium-fertility variant, which is regarded as the most likely compared to extreme low and high variants. In Table 1, we only mention countries with a distance smaller than one, and include those beyond one starting with 2060 to highlight countries similar today to the world of the distant future. Figure 3 depicts the similarity of the world population to each country every ten years using the world map, and Figure 4 shows the population pyramids of the world at each period and the most similar country in 2015. Figure 5 shows the transition of the Aitchison distance for countries most similar to the world throughout the years. At each point where the Aitchison distance is the smallest, the corresponding country depicts the population composition ratio of the world at that point in time. Next, we examine each period and the countries and regions that epitomize the world and elaborate the results.

---

[1] For detail on the aggregation of regions with populations of fewer than 90,000 see United Nations (2015).

**[Figure 2: Aitchison distance and similarity of population pyramid shapes], around here**

**[Table 1: Areas, countries, and regions most similar to the past and future world in terms of population structure based on the Aitchison distance], around here**

**[Figure 3: Degree of similarity of the World population to each country in 2015**

**throughout the years], around here**

**[Figure 4: Population pyramids (percent) of the world at each period (Top) and the corresponding most similar country in 2015 (Bottom)], around here**

**[Figure 5: Transition of the Aitchison distance for countries most similar to the world throughout the years], around here**

### 3.1. The epitome of 1990 and 2000

India epitomizes the world population structure in 1990 and Algeria in 2000.

71 percent of the population in Southern Asia lives in India, making its population structure the most dominant and yielding the results included in Table 1. Next is Pakistan with 10 percent and Bangladesh with 8 percent (United Nations, 2015). Bangladesh appears in the similarity table for 2000, although Pakistan does not, because its population structure is dissimilar to the world population of the 1990s.

Egypt and Algeria rank second and third respectively in the similarity table. The demographics of North Africa indicate that as a young region similar to Southern Asia, it enjoys the benefit of the demographic dividend and the old age dependency ratio is very low (8.4 in 2015; 14.8 in the United States).

**3.2. The epitome of 2010, 2015, 2020, and 2030**

Countries most similar to the world in 2010 and 2015 are all Latin American countries excepting Sri Lanka and Algeria. In terms of development and income, the similarity improved to match upper-middle-income countries and less developed regions, excluding the least developed regions, indicating a slight improvement in the dominant economic situation in the world starting in 2010.

In terms of the population structure in Latin America, the young population, which increased over the past century, is expected to decrease in the next fifty years. The adult population increased in the first period, and is expected to continue increasing in the next period. However, the older population is rapidly increasing, and is expected to quadruple between 2005 and 2050, when it will outnumber the young population by 2050 (Saad, 2011). For 2030, Argentina is the most similar country. Compared to most of South America, Argentina reached stage 4 of the demographic transition (birth rate and death rate are both low) earlier. This is due to a strong economy and social mobility combined with technological and medical advances.

**3.3. The epitome of 2040**

In this period, New Zealand, the United States, and Australia are the most similar countries. As a developed country, New Zealand has a relatively high fertility rate of 2.05 for the 2010–2015 period. New Zealand, similar to other developed countries, underwent demographic transition, but never experienced a natural decrease, making a natural increase the main factor in population growth. On the other hand, migration is unstable (a negative net migration rate in 1970–1990), and added to age-selective emigration at the subnational level, might be the driver of its structural aging, rather than the conventional factor, namely low fertility.

Since 2005, net migration has been the main driver of population growth in Australia, rather than a natural increase. The non-working age population (mainly those aged more than 65 years old) has been growing faster than the working age population over the past twenty years. In the last five years, the growth rate of the former is 2.3 percent, whereas the latter only grew 1.2 percent (Australian Bureau of Statistics, 2015).

The United States has doubled in size since 1950, while its industrialized counterparts have experienced slow growth during the same period (Germany: 16 percent, Italy: 28 percent). In addition to societal adaptations such as better access to child care and more involvement of males in the household, this is due to the differential fertility rates for ethnic and racial groups (Hispanic: 2.9, non-Hispanic white: 2.0 in 2008). Another major factor in population growth is net migration, projected to remain positive for the full 1950–2050 period.

### 3.4. The epitome of 2050

24 percent of Europe's population is aged more than 60 years, making it the oldest region in the world currently. (United Nations, 2015). Based on this, the epitome of the world in 2050 is likely a developed country, especially one in Europe. Indeed, six similar countries are in Europe, Northern America, and Oceania. More developed regions also made the list as an epitome of the world in 2050. However, at the top of the list are Uruguay and Puerto Rico, which belong to the Latin American and the Caribbean region and are classified as less-developed countries.

These two countries do not seem to be an appropriate fit as epitomes of the world in 2050. However, their population structure indicates that immigration issues and history can explain their similarity to the world of 2050.

The demographic structure and growth of Uruguay is distinct from its Latin American context, and similar to that of Western Europe. One reason is the cultural influence of the early intense European immigration on the scattered population, which marked an early adaptation of the European social model. (In 1950, the total fertility rate of Europe and Uruguay was 2.7, and 5.9 in Latin America and the Caribbean.) Furthermore, intense urbanization prevented the development of rural areas where reproduction levels would be high. Adding to the latter, an authoritarian regime that ended in 1984 followed by slow democratic reform resulted in an intense emigration movement. Age-selective emigration and emigration in blocks (emigration of the entire family) have decreased the juvenile and working population. Coupled with the low fertility rate and high life expectancy, the aging population has become a significant issue for policy makers in Uruguay.

Similarly, Puerto Rico's population is shrinking. The natural increase in population is one reason for this loss. The birth rate has been declining, and an aging population has increased the death rate, leading to a negative natural population increase over the past four decades. Granting the people of Puerto Rico with U.S. citizenship in 1917, and the introduction of low-cost flights have improved the freedom of movement between the island and the U.S. mainland, making emigration -mainly of those aged 16–30 years- a major driver of the sharp decline in population. and of the acceleration of the aging process of the population.

"Although the populations of all countries are expected to age over the foreseeable future, the population will remain relatively young, at least in the short term, in countries where fertility is still high" (United Nations, 2015). This is a combination of the characteristics of more developed regions and less developed regions, making Uruguay a pertinent result as the epitome of the world in 2015 as it is a developing country following a European demographic model.

### 3.5. The epitome of 2060 and beyond

Beyond 2060, the number of similar countries declines again, and Puerto Rico remains at the lead. However, from 2080, Japan takes the lead followed by Italy, the two countries that today suffer most from shrinking populations and rapidly expanding aging populations. The population aged 60 years and more is the fastest growing and expected to increase to 3.1 billion by 2100 (United Nations, 2015). However, when considering changes in life expectancy, the speed of aging—likely to accelerate in the upcoming decades—will decelerate by 2050 (Lutz et al., 2008). Forecasting the distant future is a challenging task yielding mostly uncertain results, since it requires a preliminary forecast of the near future, which is also pending.

### 3.6. A comprehensive epitome: Cluster analysis

We conducted a cluster analysis to summarize the findings. We employ the Ward method based on the calculation algorithm proposed by Murtagh and Legendre (2014). Figure 6 shows the clusters of countries and regions in 2015 based on the similarity of their population composition ratios. The distribution of the lower value of the Aitchison distance indicates the group of countries and regions that are similar, and helps link each cluster to a specific period: past, present, near future, and distant future.

**[Figure 6: Cluster map of countries and regions based on the similarity of their population composition ratios in 2015 (Top) and corresponding population pyramids (percent) (Bottom)], around here**

From the generated cluster map, we identified seven types of clusters. Clusters 1, 2, and 3 have a population structure with a pyramid shape, characterized by both high birth and mortality rates. This is characteristic of Sub-Saharan Africa, the Middle East, South-Eastern Asia, and a few countries in southern Asia including Afghanistan and Pakistan. It represents the trend that was most dominant of the world population structure in the past (before 1990).

Cluster 2 was identified as a sub-cluster of Cluster 1, with a relatively high population. These clusters have long suffered a high mortality rate in the past decades, but also have a large productive population (high fertility rate), which has resulted in a large juvenile population (United Nations, 2015).

The population structure of Cluster 4 has a bell shape. It is characterized by high birth and mortality rates and a stagnant population. Representative areas are China, North Africa, Southern Asia, Brazil, India, Central Asia, and South Africa. While these areas have a high birth rate, they face a high mortality rate (United Nations, 2015). This cluster represents the world at the beginning of the century.

The population structure of Cluster 5 has a spindle shape. It is characterized by a declining juvenile population and growing aging population, a low birth rate, and a declining population. Representative areas are Latin America and Israel. These countries are not yet in the phase where the aging population is dominant; however, the trend cautions for that risk. This cluster represents the world in the present and the near future.

The population structure of Cluster 6 has a star shape. It is characterized by a small juvenile population, a small aging population, and a large-working population. Representative areas are Russia and Western Europe. Furthermore, this cluster does not represent the world at any period, and represents a recessive population structure. The cluster is classified by the UN as economies in transition, which might clarify the peculiar population structure of the cluster.

The population structure of Cluster 7 has a coffin shape. It is characterized by drastically declining birth and population rates, and extreme growth of the aging population. Representative areas are Japan, Europe, North America, and Oceania. The world is heading towards this structure in the distant future.

## 4. Conclusions

Focusing on the total size of the population or on one particular age group is not sufficient to comprehensively forecast demographic changes. At the same time, studying all age groups simultaneously through multiple periods is a challenging task. Therefore, we took advantage of CoDA to examine the transition of the world population as a structure from 1990 to the end of the 21$^{st}$ century by comparing the world population structure of each period to that of countries and regions in 2015. We found that in 1990, the world was most similar to India, Algeria and Egypt, which clarifies that the demographic characteristics of the world at that time were a relatively high fertility rate, small aging population, and large juvenile and working population. The world has not yet achieved demographic transition. In the present period and near future (2015–2030), the world is most similar to Latin American and the Caribbean countries, where the young population is still relatively high, although the trend is converging towards an aging population. In 2040, the world is most similar to developed countries (New Zealand, the United States, and Australia), where the population is still growing because of a natural increase and net migration, and where the aging population is relatively high, but not as high as in Japan or Italy. In 2050, something interesting occurs. Where developed countries were expected to appear, especially those in aging Europe, the most similar countries were Uruguay and Puerto Rico. In these countries, the population is shrinking and aging, not because of a natural decrease, but because of intensive age-selective migration and the strong early European influence in the case of Uruguay. In 2060 and beyond, the similarity of the world to other countries becomes difficult to ascertain, as the dissimilarity of the shapes of the pyramids is remarkable when the Aitchison distance is higher than one. However, in order of similarity, Japan and Italy emerge at the top of the list, confirming the forecast of a stagnating or shrinking world population in the distant future, with an all-time high rate of the aging population.

In this research, based on the characteristics of population composition ratio data, an analysis was conducted using the Aitchison distance, a basic distance index in CoDA. However, as Otero et al. (2005) point out, analysts select the distance indicator based on what they intend to emphasize according to the subject at hand. At present, in population research and regional analysis, application examples of the Aitchison distance are limited, and further discussion about the suitability of its application is necessary.

**Figures**

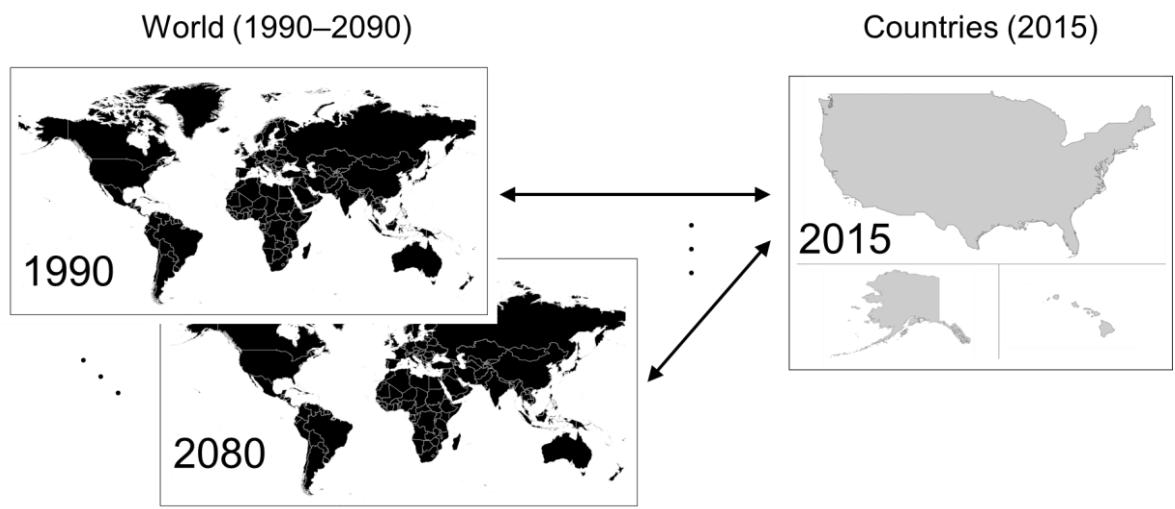

**Figure 1**: Analysis scheme: Comparison between world population structures of countries and areas in 2015 and different points of time

| Class | World 2015 | Colombia 2015 | Sri Lanka 2015 | Brazil 2015 | Thailand 2015 | Pakistan 2015 |
|---|---|---|---|---|---|---|
| 100+ | 0.01 | 0.01 | 0.00 | 0.00 | 0.01 | 0.00 |
| 95–99 | 0.04 | 0.04 | 0.03 | 0.03 | 0.05 | 0.01 |
| 90–94 | 0.18 | 0.16 | 0.16 | 0.14 | 0.20 | 0.04 |
| 85–89 | 0.50 | 0.39 | 0.46 | 0.45 | 0.61 | 0.17 |
| 80–84 | 0.97 | 0.74 | 0.87 | 0.86 | 1.24 | 0.43 |
| 75–79 | 1.56 | 1.19 | 1.35 | 1.47 | 2.03 | 0.83 |
| 70–74 | 2.08 | 1.79 | 2.46 | 2.01 | 2.62 | 1.31 |
| 65–69 | 2.93 | 2.73 | 3.97 | 2.87 | 3.72 | 1.71 |
| 60–64 | 3.98 | 3.80 | 4.63 | 3.89 | 5.32 | 2.11 |
| 55–59 | 4.62 | 4.74 | 5.57 | 4.86 | 6.71 | 2.95 |
| 50–54 | 5.47 | 5.82 | 6.08 | 5.94 | 7.62 | 3.75 |
| 45–49 | 6.18 | 6.63 | 6.39 | 6.35 | 8.28 | 4.39 |
| 40–44 | 6.60 | 6.54 | 6.59 | 6.89 | 8.32 | 5.06 |
| 35–39 | 6.77 | 7.37 | 7.49 | 7.84 | 8.13 | 6.08 |
| 30–34 | 7.50 | 8.15 | 7.32 | 8.57 | 7.40 | 7.35 |
| 25–29 | 8.30 | 8.49 | 6.99 | 8.37 | 6.84 | 8.89 |
| 20–24 | 8.21 | 8.70 | 7.41 | 8.02 | 6.71 | 9.70 |
| 15–19 | 8.03 | 8.44 | 7.65 | 8.40 | 6.48 | 10.22 |
| 10–14 | 8.26 | 8.44 | 8.20 | 8.38 | 6.18 | 10.32 |
| 05–09 | 8.67 | 8.10 | 8.44 | 7.41 | 5.94 | 11.64 |
| 00–04 | 9.13 | 7.75 | 7.93 | 7.23 | 5.59 | 13.05 |
| Aitchison distance between World 2015 and country 2015 | | 0.532 | 0.639 | 0.900 | 1.152 | 3.309 |

**Figure 2:** Aitchison distance and similarity of population pyramid shapes

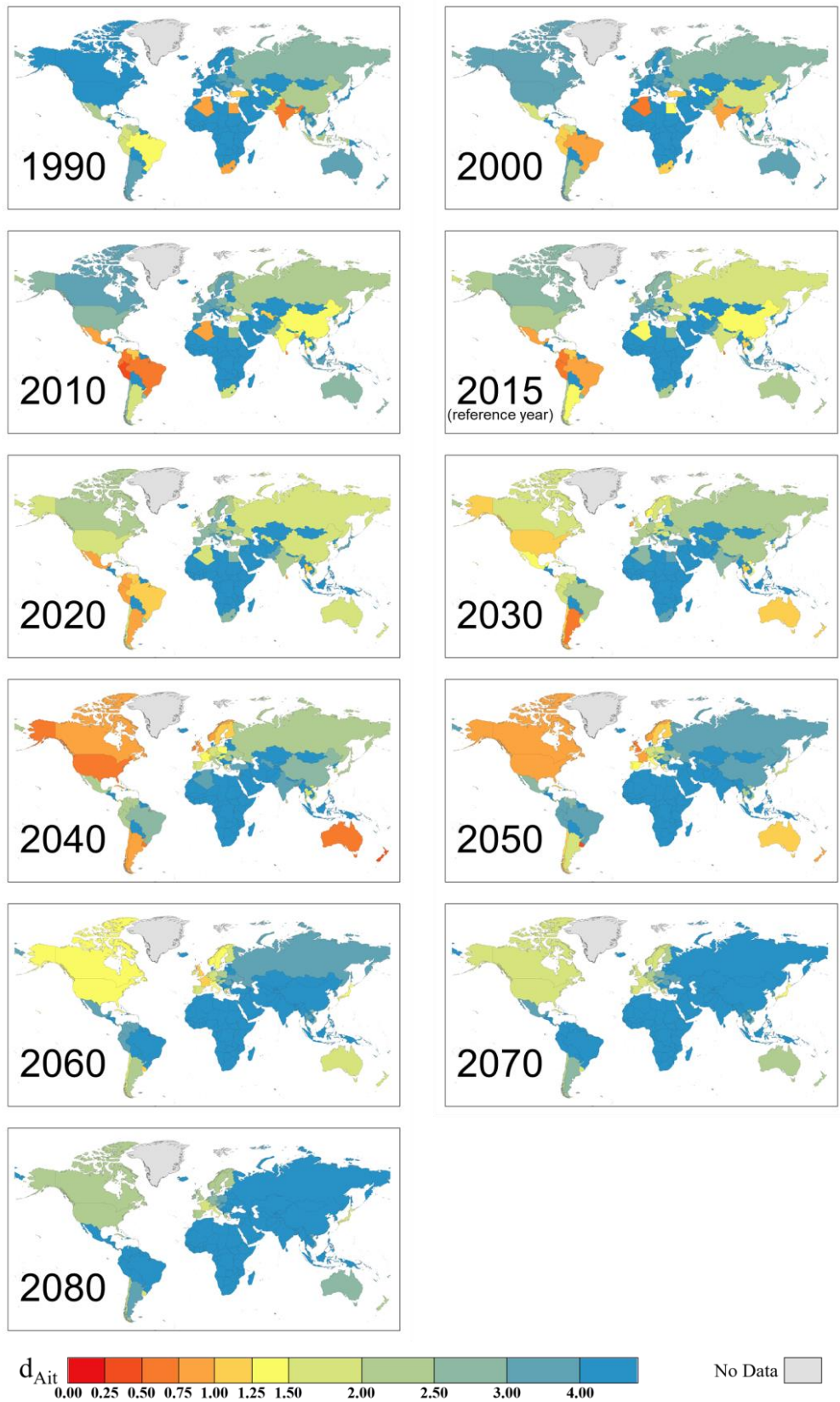

**Figure 3**: Degree of similarity of the world population of each country in 2015 throughout the years

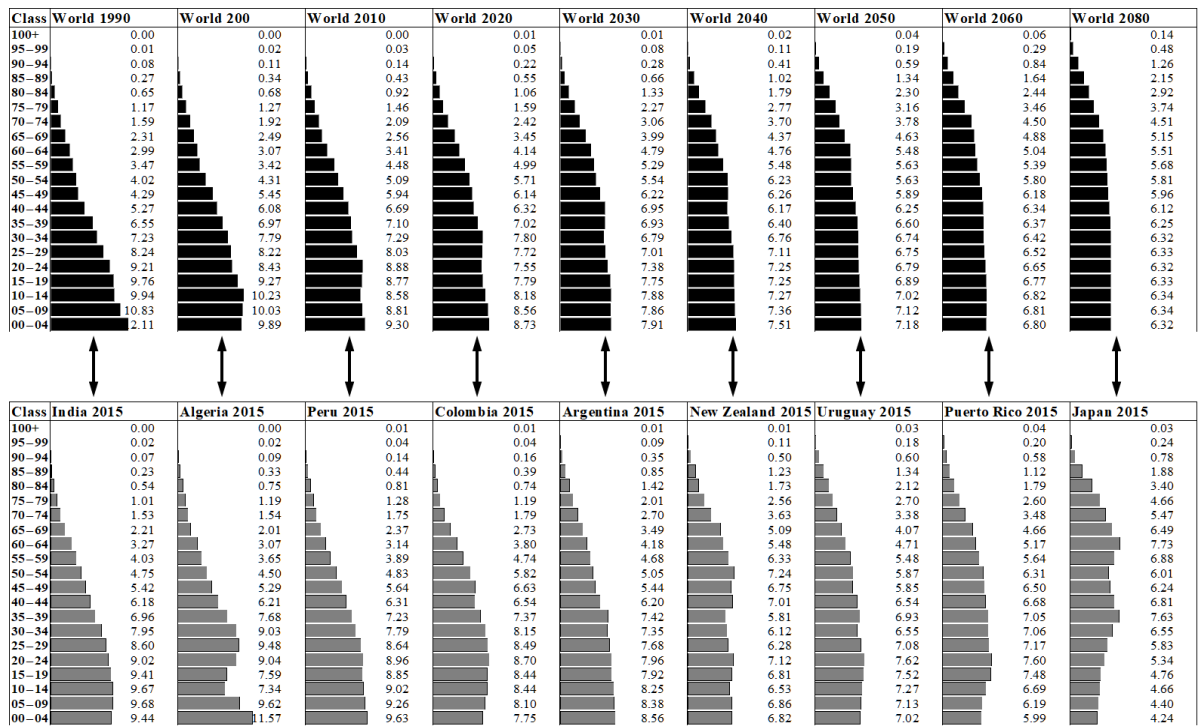

**Figure 4**: Population pyramids (percent) of the world at each period (Top) and the corresponding most similar country in 2015 (Bottom)

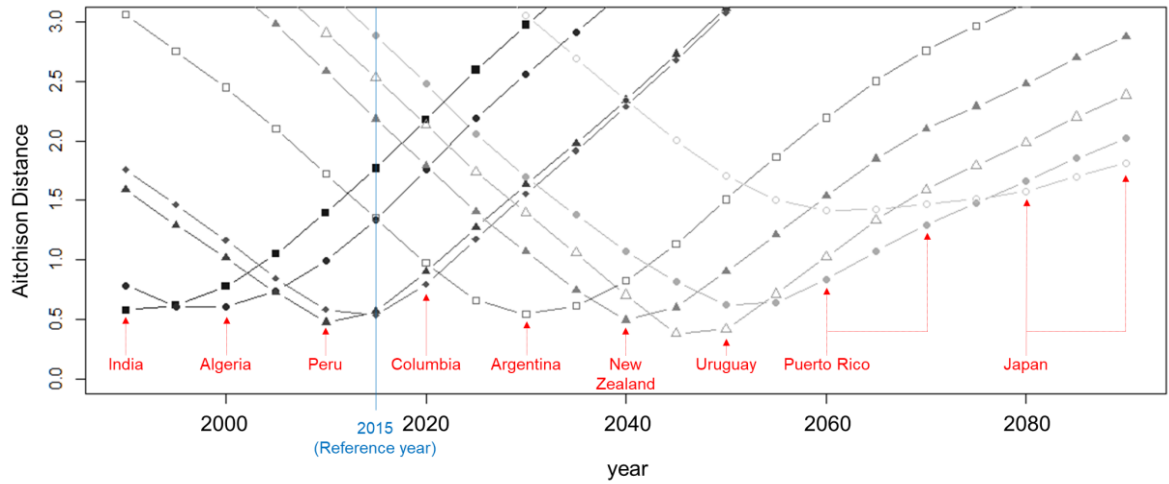

**Figure 5**: Transition of the Aitchison distance for countries most similar to the world throughout the years

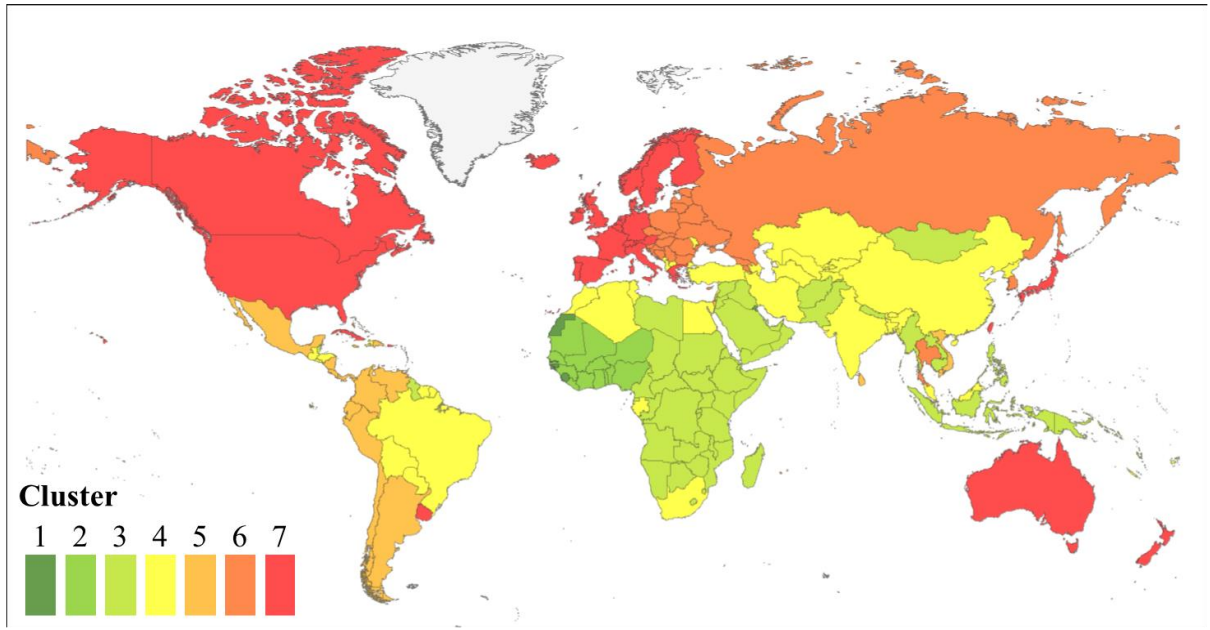

| Class | Cluster 1 | Cluster 2 | Cluster 3 | Cluster 4 | Cluster 5 | Cluster 6 | Cluster 7 |
|---|---|---|---|---|---|---|---|
| 100+ | 0.00 | 0.00 | 0.00 | 0.00 | 0.01 | 0.01 | 0.03 |
| 95–99 | 0.00 | 0.00 | 0.00 | 0.02 | 0.04 | 0.07 | 0.17 |
| 90–94 | 0.01 | 0.01 | 0.03 | 0.10 | 0.29 | 0.23 | 0.71 |
| 85–89 | 0.05 | 0.04 | 0.14 | 0.33 | 1.02 | 0.53 | 1.62 |
| 80–84 | 0.18 | 0.17 | 0.36 | 0.77 | 1.86 | 0.93 | 2.55 |
| 75–79 | 0.45 | 0.39 | 0.70 | 1.34 | 3.20 | 1.33 | 3.44 |
| 70–74 | 0.84 | 0.74 | 1.11 | 1.90 | 3.20 | 1.86 | 4.28 |
| 65–69 | 1.27 | 1.18 | 1.58 | 2.79 | 4.42 | 2.55 | 5.55 |
| 60–64 | 1.72 | 1.71 | 2.19 | 4.13 | 6.44 | 3.58 | 6.02 |
| 55–59 | 2.28 | 2.53 | 2.88 | 4.69 | 7.21 | 4.50 | 6.67 |
| 50–54 | 2.86 | 3.28 | 3.58 | 5.76 | 7.37 | 5.33 | 7.25 |
| 45–49 | 3.53 | 4.46 | 4.33 | 6.83 | 6.84 | 6.05 | 7.08 |
| 40–44 | 4.42 | 5.70 | 5.14 | 7.14 | 7.43 | 6.64 | 6.93 |
| 35–39 | 5.48 | 7.28 | 6.05 | 7.03 | 7.67 | 7.31 | 6.45 |
| 30–34 | 6.55 | 8.76 | 7.16 | 7.84 | 7.80 | 7.84 | 6.56 |
| 25–29 | 7.58 | 8.92 | 8.03 | 9.03 | 7.66 | 8.40 | 6.33 |
| 20–24 | 8.80 | 8.79 | 9.14 | 8.49 | 6.37 | 8.74 | 6.22 |
| 15–19 | 10.39 | 9.32 | 10.31 | 7.85 | 5.23 | 8.46 | 5.70 |
| 10–14 | 12.31 | 10.49 | 11.29 | 7.84 | 4.98 | 8.44 | 5.52 |
| 05–09 | 14.41 | 12.31 | 12.34 | 7.98 | 5.37 | 8.55 | 5.54 |
| 00–04 | 16.87 | 13.92 | 13.63 | 8.16 | 5.60 | 8.66 | 5.38 |

**Figure 6**: Cluster map of countries and regions based on the similarity of their population composition ratios in 2015 (Top) and corresponding population pyramids (percent) (Bottom)

**Table 1:** Countries and regions most similar to the past and future world in terms of population structure based on the Aitchison distance

| Rank | 1990 | | 2000 | | 2010 | | 2015 (as reference year) | | 2020 | | 2030 | | 2040 | | 2050 | | 2060 | | 2070 | | 2080 | |
|---|---|---|---|---|---|---|---|---|---|---|---|---|---|---|---|---|---|---|---|---|---|---|
| | Country | $d_{Ait}$ | Country | $d_{Ait}$ | Country | $d_{Ait}$ | Country | $d_{Ait}$ | Country | $d_{Ait}$ | Country | $d_{Ait}$ | Country | $d_{Ait}$ | Country | $d_{Ait}$ | Country | $d_{Ait}$ | Country | $d_{Ait}$ | Country | $d_{Ait}$ |
| *Countries* | | | | | | | | | | | | | | | | | | | | | | |
| 1 | India | 0.579 | Algeria | 0.605 | Peru | 0.474 | Colombia | 0.532 | Colombia | 0.792 | Argentina | 0.544 | New Zealand | 0.494 | Uruguay | 0.419 | Puerto Rico | 0.839 | Puerto Rico† | 1.293 | Japan† | 1.576 |
| 2 | Egypt | 0.776 | India | 0.775 | Ecuador | 0.581 | Peru | 0.567 | Sri Lanka | 0.801 | Ireland | 0.776 | United States | 0.635 | Puerto Rico | 0.622 | Uruguay† | 1.027 | Japan† | 1.469 | Puerto Rico† | 1.662 |
| 3 | Algeria | 0.785 | Brazil | 0.847 | Colombia | 0.581 | Ecuador | 0.594 | Ecuador | 0.838 | Israel | 0.789 | Australia | 0.655 | United Kingdom | 0.732 | France† | 1.125 | Italy† | 1.501 | Italy† | 1.697 |
| 4 | | | Bangladesh | 0.877 | Brazil | 0.694 | Sri Lanka | 0.639 | Mexico | 0.872 | | | Ireland | 0.677 | United States | 0.760 | United Kinddom† | 1.149 | France† | 1.556 | France† | 1.858 |
| 5 | | | | | Sri Lanka | 0.706 | Mexico | 0.806 | Peru | 0.900 | | | Uruguay | 0.703 | Norway | 0.882 | Italy† | 1.321 | Uruguay† | 1.589 | Uruguay† | 1.983 |
| 6 | | | | | Mexico | 0.939 | Brazil | 0.900 | Argentina | 0.975 | | | Chile | 0.820 | New Zealand | 0.900 | | | | | | |
| 7 | | | | | Algeria | 0.989 | | | | | | | Argentina | 0.824 | France | 0.916 | | | | | | |
| 8 | | | | | | | | | | | | | Norway | 0.863 | Canada | 0.926 | | | | | | |
| 9 | | | | | | | | | | | | | Canada | 0.956 | | | | | | | | |
| 10 | | | | | | | | | | | | | United Kingdom | 0.960 | | | | | | | | |
| 11 | | | | | | | | | | | | | Israel | 0.965 | | | | | | | | |
| 12 | | | | | | | | | | | | | Cuba | 0.996 | | | | | | | | |
| *Geographic regions* | | | | | | | | | | | | | | | | | | | | | | |
| 1 | Southern Asia | 0.530 | Southern Asia | 0.812 | Asia | 0.449 | South America | 0.447 | Oceania | 0.662 | Northern America | 0.661 | Northern America | 0.767 | | | | | | | | |
| 2 | Western Asia | 0.606 | Asia | 0.841 | South America | 0.590 | Latin America and Caribbean | 0.502 | Caribbean | 0.719 | Oceania | 0.891 | Northern Europe | 0.860 | | | | | | | | |
| 3 | Southern Africa | 0.792 | South-Eastern Asia | 0.888 | Latin America and Caribbean | 0.715 | Central America | 0.935 | | | Northern Europe | 0.947 | | | | | | | | | | |
| 4 | Northern Africa | 0.813 | Central Asia | 0.944 | South-Eastern Asia | 0.826 | Oceania | 0.968 | | | | | | | | | | | | | | |
| 5 | Central Asia | 0.904 | Western Asia | 0.988 | Central America | 0.882 | | | | | | | | | | | | | | | | |
| *Regions categorized by UN* | | | | | | | | | | | | | | | | | | | | | | |
| 1 | Less developed (excluding China) | 0.640 | Less developed | 0.497 | Less developed (excluding least developed) | 0.671 | Less developed (excluding least developed) | 0.980 | | | | | | | | | More developed | 0.961 | | | | |
| 2 | Less developed | 0.810 | Less developed (excluding China) | 0.624 | Less developed | 0.847 | | | | | | | | | | | | | | | | |
| 3 | | | Less developed (excluding least developed) | 0.629 | | | | | | | | | | | | | | | | | | |
| *Regions categorized by World Bank* | | | | | | | | | | | | | | | | | | | | | | |
| 1 | Lower-middle-income | 0.561 | Middle-income | 0.561 | Middle-income | 0.804 | Upper-middle-income | 0.902 | | | | | High-income | 0.808 | High-income | 0.877 | | | | | | |
| 2 | Middle-income | 0.916 | Lower-middle-income | 0.790 | Upper-middle-income | 0.854 | | | | | | | | | | | | | | | | |

Note: the symbol † refers to the countries with distance over 1.0 to give an idea about the distant future, otherwise we only have one country (Puerto Rico) in 2060 and zero countries in 2070, and thus won't be able to discuss the epitome of the world of the distant future.